\documentclass[]{spie}  %>>> use for US letter paper
\usepackage{amsmath,amsfonts,amssymb}
\usepackage{graphicx}
\usepackage{epstopdf}
\usepackage[colorlinks=true, allcolors=blue]{hyperref}

\title{Private key and password protection by steganographic image encryption}

\author[a]{Debesh Choudhury}
\author[b]{Sujoy Chakraborty}
\affil[a]{Infosensys Research and Engineering, Kolkata, WB 700110, India}
\affil[b]{Stockton University, Galloway, NJ 08205, USA}
\authorinfo{Further author information: E-mail: debesh@iitbombay.org, Sujoy.Chakraborty@stockton.edu}

\begin{document} 
\maketitle

\begin{abstract}
We propose a technique to protect and preserve a private key or a passcode in an encrypted two-dimensional graphical image. The plaintext private key or the passcode is converted into an encrypted QR code and embedded into a real-life color image with a steganographic scheme. The private key or the passcode is recovered from the stego color image by first extracting the encrypted QR code from the color image, followed by decryption of the QR code. The cryptographic key for encryption of the QR code is generated from the output of a Linear Feedback Shift Register (LFSR), initialized by a seed image chosen by the user. The user can store the seed image securely, without the knowledge of an attacker. Even if an active attacker modifies the seed image (without knowledge of the fact that it is the seed image), the user can easily restore it if he/she keeps multiple copies of it, so that the encryption key can be regenerated easily. Our experiments prove the feasibility of the technique using sample private key data and real-life color images.  
\end{abstract}


\section{INTRODUCTION}
\label{sec:intro}  % \label{} allows reference to this section

Public key cryptography is widely used for diverse consumer applications ranging from digital signatures, time stamping, secure network communication, cryptocurrency, data protection, etc.~\cite{RSA,crypto:book}. A private key is a necessary credential in public-key cryptography. If the private key is lost, all the digital assets associated with it is lost forever. This is a chronic problem of public-key cryptography. This is known as the private key-loss conundrum~\cite{conundrum}. Preservation of the private key is a challenge because it is prone to stealing, hacking, or loosing. Secret preservation of the private keys may serve as a solution.

Preserving the private keys and complex text passcodes is extremely challenging. One can’t store it in the cloud, nor can write it down on paper and store it. There is a vulnerability of losing the same, or it can even be stolen. This research paper aims to develop a technique to offer a solution to the private key-loss conundrum of public-key cryptography. We plan to secretly preserve the private key as encrypted data inside a color image of our choice, such as a memorable image connected to our episodic life. A convenient mechanism is required so that the private key can be re-generated from the encrypted image. Forgetting an episodic event or object is not easy~\cite{episodic:memory}. Thus, it is hard to forget the image and lose the hidden private key inside the episodic image. It is also hard to guess that an image can contain a private key. The plaintext private key will remain safe and hidden inside a real-life color image. We keep the encryption mechanism robust so that the private key can’t easily be stolen even someone guesses/knows about the image-based private key protection.

\section{METHODOLOGY}
\label{sec:method}

In this paper we propose to protect and preserve the plaintext private key(s) inside episodic color images connected to our life~\cite{episodic:memory}. (i)First we convert the plaintext private key into a QR code image. (ii) Then we encrypt the QR code image into a random white noise image by a novel image encryption technique using a visual password image, that would generate the encryption key for a stream cipher from the output of a Linear Feedback Shift Register (LFSR)~\cite{LFSR:2021}. (iii) Finally, the white noise encrypted image is concealed inside a real-life color image using a strong steganographic algorithm~\cite{stego}. The plaintext private key can be recovered by carrying out the reverse operation of the steps (iii), (ii) and (i), i.e., de-conceal, decrypt and decode QR coded plaintext. We describe each step in the following subsections.
   \begin{figure} [ht]
   \begin{center}
   \includegraphics[width=4cm, angle=-90]
   {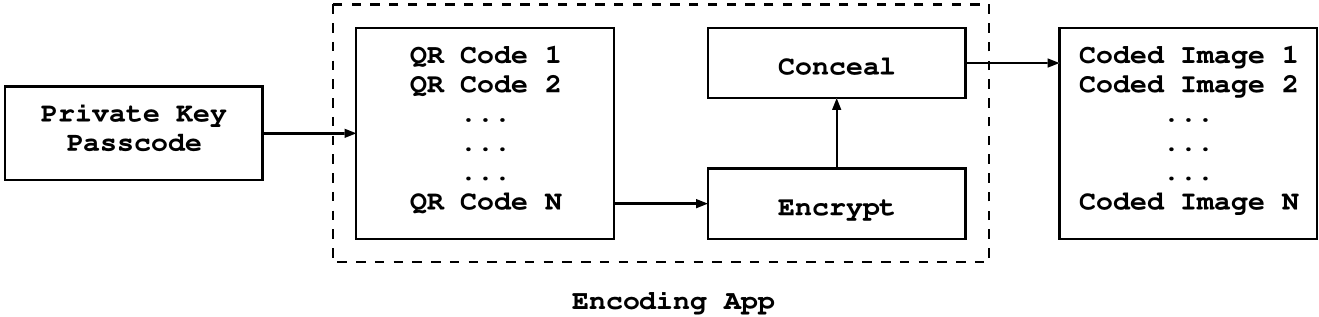}
   %{flowchart_image_pass_encode.eps}
   \\[4pt]
   (a)\\[7pt]
   \includegraphics[width=4cm, angle=-90]
   {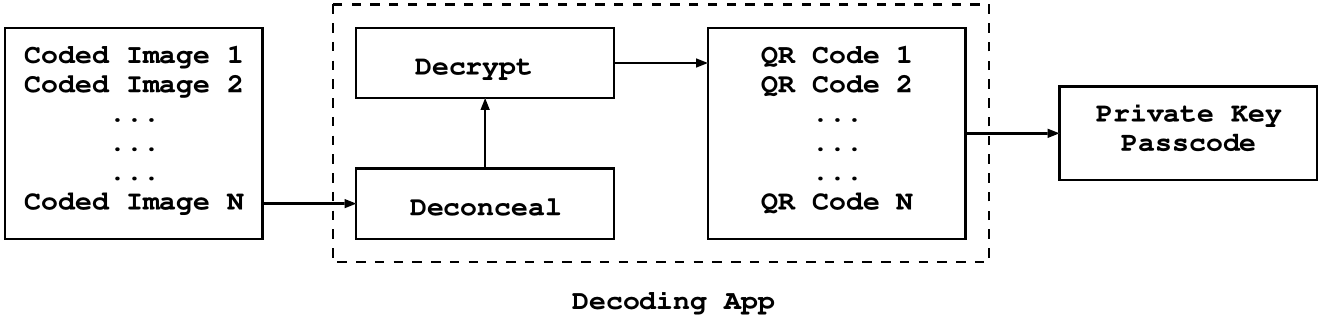}
   %{flowchart_image_pass_decode.eps}
   \\[4pt]
   (b)
   \end{center}
   \caption[example] 
   { \label{fig:flowchart} 
   The flowcharts of the (a) encoding algorithm, and (b) decoding algorithm.}
   \end{figure} 

\subsection{Converting plaintext private key to QR code}The plaintext private key is first converted to QR code. We used standard open source Linux application software for conversion of plaintext to QR code~\cite{QRencode:Linux}.
\subsection{Encryption of the QR code}To encrypt the QR code, we use an image that acts as a visual password to generate the encryption key. The benefit of using an image as a visual password is that the user doesn't have to remember or store the encryption key. The encryption key is automatically generated from the output of a Linear Feedback Shift Register (LFSR)~\cite{LFSR:2021}. The output of LFSR has excellent stochastic property of randomness and thus serves as a strong encryption key. The steps of encryption of the QR code are as follows:
\subsubsection{Choose a color image as a visual password}The first step is to choose a color image that will generate the encryption key. The benefit of using a visual password is that the user doesn't have to remember/store the password in its original form, but just need to remember the image that generates the same.
\subsubsection{Separate the three color channels}The next step is to separate the three color channels from the visual password image and store them as individual grayscale images. The goal is to generate binary images from individual color channels to create a bit stream that would set the initial state of a Linear Feedback Shift Register (LFSR).
\subsubsection{Apply thresholding to each color channel}The next step is to threshold the grayscale image corresponding to each color channel obtained from the previous step to the intensity value of $128$. This creates three binary images corresponding to the individual color channels (pixels with intensity greater than or equal to 128 are bit $1$ and pixels with intensity less than 128 are bit $0$). 
\subsubsection{Generate a binary bit stream from the binary images}We generate a stream of bits from the three binary images to initialize the state of a $128$-bit LFSR. To generate the $128$ bits, we consider pixels row-wise from the binary images obtained from the three color channels. The first bit is chosen from the binary image of the Red channel, the second bit is chosen from the Green channel, the third bit from the Blue channel, fourth from the Red channel, fifth from the Green channel and so on. This creates a binary bit pattern.
   \begin{figure} [ht]
   \begin{center}
   \includegraphics[width=\textwidth]
   {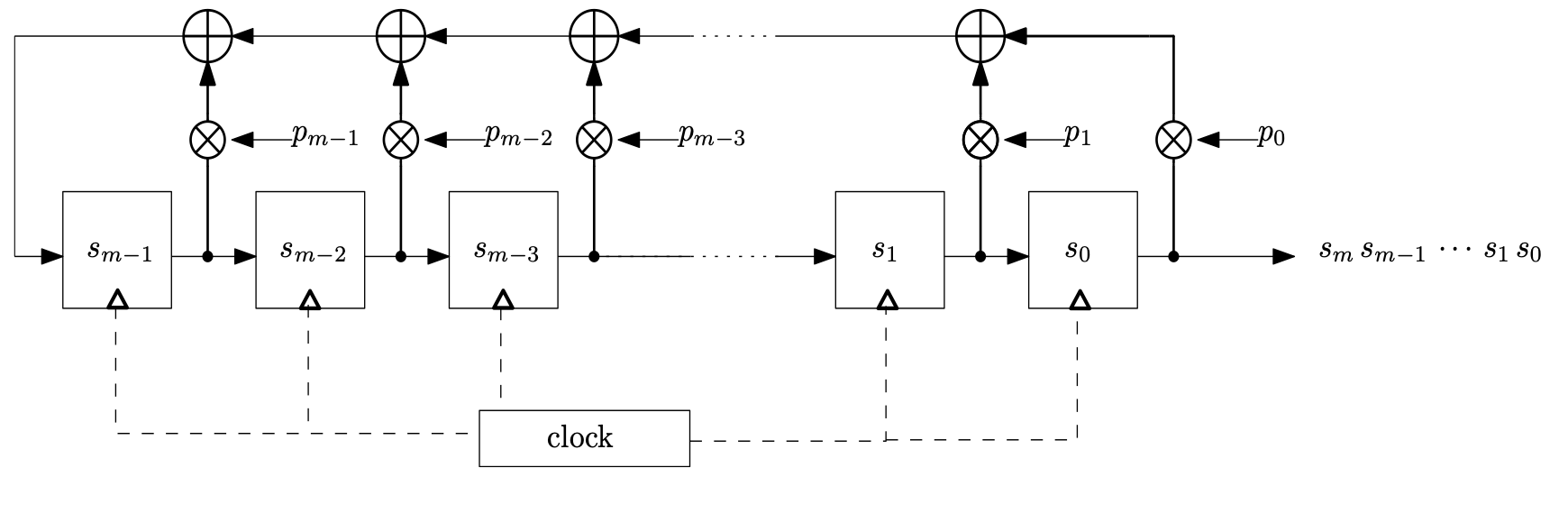}
   \caption{A General LFSR of $m$ bits. Adopted from [\citenum{LFSR:2021}].}
   \label{fig:LFSR}
   \end{center}
\end{figure}    
\subsubsection{Initialize the LFSR}The first $128$ bits of the generated bit stream is then used to set the initial state of a $128$-bit LFSR. The output of the LFSR is then used to encrypt the QR code, which is also a binary image.
\subsubsection{Encrypting the QR code}The QR code is encrypted with the $128$-bit binary output of the LFSR~\cite{LFSR:2021}. Let's say the QR code is of size $M \times N$. First, the QR code image is converted to a column vector, where we take each bit from the QR code image column-wise. Then the first M * N bits of the LFSR output is XOR-ed with the column vector bit-wise, which encrypts the QR code. The encryption operation in this case is the simple XOR operation. 
\subsection{Concealing the encrypted image}The encrypted QR code is then concealed in a cover image using a strong steganographic scheme, in which we hide the encrypted QR code in the Blue channel of the cover image~\cite{stego}. The cover image can be chosen by the user as per their preference. Thus, the entire scheme of private key protection is driven only by the selection of the visual password image and the cover image by the user, and everything else follows automatically.  

Figure~\ref{fig:LFSR} shows a general Linear Feedback Shift Register (LFSR) of $m$ bits. It is important to note here that for an LFSR of $m$ bits with maximum sequence length, we have a sequence length of $2^m - 1$. Thus, for a $128$-bit LFSR, we have an output sequence length of $2^{128} - 1$. It is impossible to launch a brute-force attack against a $128$-bit LFSR, which ensures the security of this encryption scheme. Also, it is almost impossible to guess that private key(s) are preserved inside the real-life color image(s) as encrypted QR code data. Moreover, the data are protected by two-stage cryptographic processing. The chances of guessing and breaking the key(s) are extremely low. Since the private key data are preserved inside real-life color images of the users, which are most likely connected to the users' notable episodic past, the users' images or photos are easily remembered. However, it will be hard to guess and crack by the hackers.

\section{EXPERIMENTS}

The flowchart of the encoding and the decoding algorithms are shown in Fig.\ref{fig:flowchart}(a) and Fig.\ref{fig:flowchart}(b) respectively. The encoding algorithm of Fig.\ref{fig:flowchart}(a) performs the processes described in steps (i), (ii) and (iii) on the private keys or the passcodes. The decoding algorithm of Fig.\ref{fig:flowchart}(b) performs the processing steps in the reverse order, i.e., (iii), (ii) and (i) on the encoded images.

\begin{figure} [ht]
   \begin{center} 
   E9873D79C6D87DC0FB6A5778633389F4453213303DA61F20BD67FC233AA33262
   \\[5pt] (a)\\[10pt]
   \includegraphics[height=4.75cm]
   {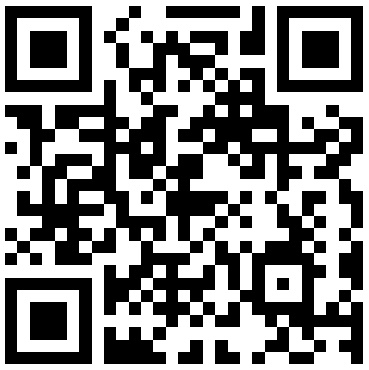}
   \hspace{3.75cm}
   \includegraphics[height=4.75cm]
   {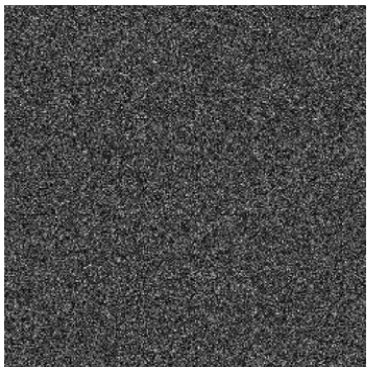}
   \\[5pt]
   (b) \hspace{8cm} (c)\\[10pt]
   \end{center}
   \hspace{0.8cm}
   \includegraphics[height=4.75cm]
   {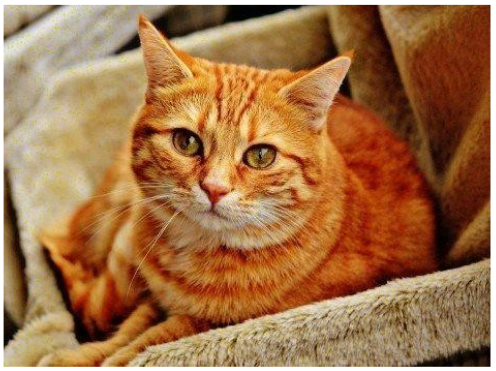}
   \hspace{3cm}
   \includegraphics[height=4.75cm]
   {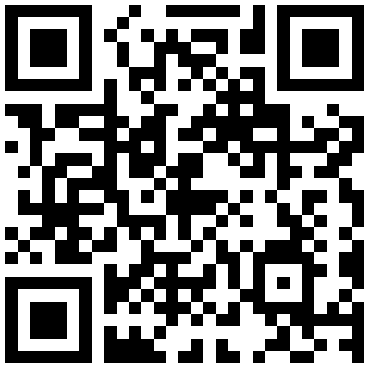}
   \begin{center}
   (d) \hspace{8cm} (e)\\[10pt]
   E9873D79C6D87DC0FB6A5778633389F4453213303DA61F20BD67FC233AA33262
   \\[5pt] (f)
   \end{center}
   \caption[example] 
   { \label{fig:results} 
   Experimental results: (a)~Sample private key, (b)~QR Code of the private key, (c)~Encrypted QR Code, (d)~Encrypted QR code concealed in a color image, (e)~Recovered QR code image, (f)~Recovered private key.}
   \end{figure}

The feasibility of the proposed technique is demonstrated by using an example private key~\cite{privatekey}. We have converted the plaintext private key into a QR code. The sample private key and its QR code are shown in Fig.\ref{fig:results}(a) and \ref{fig:results}(b) respectively. We have encrypted the QR code using LFSR based encryption algorithm~\cite{LFSR:2021} which is shown in Fig.\ref{fig:results}(c).  The encrypted QR code image is then concealed inside a color image of a cat using a steganographic algorithm~\cite{stego} which is shown in Fig.\ref{fig:results}(d). We 
have also recovered the plaintext private key by following the de-concealing of the encrypted QR code from the stego color image, followed by decryption and reading of the QR code~\cite{QRencode:Linux}. These are shown in Fig.\ref{fig:results}(e) and \ref{fig:results}(f).

We haven't shown the image used as a visual password to encrypt the QR code image by the LFSR technique. It can be any color image connected to users' episodic life. The approach presented here is simple, yet effective. It can be noted however, that the encryption scheme could be made further stronger to prevent possible attacks on the LFSR, by replacing the single LFSR with combinations of multiple LFSR-s (such as the Trivium\cite{Par:2009} or the A5/1 cipher\cite{A5-1:2005}). We leave it for the future work.

\section{CONCLUSION}

We have utilized simple yet strong encryption and steganographic techniques to secretly preserve QR coded plaintext private key data inside episodic color images connected to real-life. This research indirectly solves the age-old problem of public-key cryptography known as “private key-loss conundrum”~\cite{conundrum}. This technique can permanently solve the chronic problem of “private key-loss conundrum” in various real-world use cases associated with cryptocurrency and decentralized distributed ledger technologies~\cite{privatekey,bitcoin,bitcoin2,DLT}. This technique can also be utilized to preserve any long and complex text passcodes~\cite{password:book}.

\end{document}